\documentclass[12pt,preprint]{aastex}

\usepackage{graphicx}

\usepackage{natbib}

\slugcomment{ApJ., Revised: 2010-Dec-02}

\shorttitle{Comet 17P/Holmes}
\shortauthors{Li, Jewitt, Clover and Jackson}

\begin{document}


\title{Outburst of Comet 17P/Holmes Observed With The Solar Mass Ejection Imager}

\author{Jing Li\altaffilmark{1}\\
David Jewitt\altaffilmark{1,2} \\
John M. Clover\altaffilmark{3} \\
and\\
Bernard V. Jackson\altaffilmark{3}}
\affil{(1) Institute for Geophysics and Planetary Physics, University of California at Los Angeles}
\affil{(2) Depts. Earth and Space Sciences and Physics and Astronomy, University of California at Los Angeles}
\affil{(3) Center for Astrophysics and Space Science, University of California at San Diego, La Jolla, CA 92093-0424}

\email{jli@igpp.ucla.edu, jewitt@ucla.edu}

\begin{abstract}
We present time-resolved photometric observations of Jupiter family comet 17P/Holmes during its dramatic outburst of 2007.  The observations, from the orbiting Solar Mass Ejection Imager (SMEI), provide the most complete measure of the whole-coma brightness, free from the effects of instrumental saturation and with a time-resolution well-matched to the rapid brightening of the comet.  The lightcurve is divided into two distinct parts.  A rapid rise between the first SMEI observation on UT 2007 October 24 06h 37m (mid-integration) and UT 2007 October 25, is followed by a slow decline until the last SMEI observation on UT 2008 April 6 22h 16m (mid-integration).  We find that the rate of change of the brightness is reasonably well-described by a Gaussian function having a central time of UT 2007 October 24.54$\pm$0.01 and a full-width-at-half-maximum 0.44$\pm$0.02 days. The maximum rate of brightening occurs some 1.2 days after the onset of activity.  At the peak the scattering cross-section grows at 1070$\pm$40 km$^2$ s$^{-1}$ while the (model-dependent) mass loss rates inferred from the lightcurve reach a maximum at 3$\times$10$^5$ kg s$^{-1}$.   The integrated mass in the coma lies in the range (2 to 90)$\times$10$^{10}$ kg, corresponding to 0.2\% to 10\% of the nucleus mass, while the kinetic energy of the ejecta is (0.6 to 30) MTonnes TNT.   The particulate coma mass could be contained within a shell on the nucleus of thickness $\sim$1.5 to 60 m.  This is comparable to the distance travelled by conducted heat in the century since the previous outburst of 17P/Holmes.  This coincidence is consistent with, but does not prove, the idea that the outburst was triggered by the action of conducted heat, possibly through the crystallization of buried amorphous ice.  

\end{abstract}


\section{Introduction}

Comet 17P/Holmes is a dynamically unremarkable comet, with a semimajor axis, eccentricity and inclination of 3.620 AU, 0.433 and 19.1$\degr$, respectively.  The Tisserand parameter measured with respect to Jupiter is 2.86, which classifies 17P/Holmes as a member of the Jupiter comet family and suggests a likely origin in the Kuiper belt.  The perihelion distance is a modest 2.05 AU, small enough to drive the production of a coma through the sublimation of near-surface water ice but large enough that the comet is not normally spectacular as seen from the Earth.  As a result, the comet has received relatively little observational attention and the properties of its nucleus are poorly known, except for an estimate of its radius (about 1.7 km, \citealt{2009A&A...508.1045L}).  However, 17P/Holmes is distinguished by having undergone three dramatic photometric outbursts, the first leading to its discovery in November 1892 (Holmes 1892), followed by an outburst in mid-January 1893 (Barnard 1896),  and the most recent being the subject of this paper.  The recent outburst was first noticed by J. A. Henriques Santana
on UT 2007 October 24.067 \citep{2007IAUC.8886....1B} and triggered intensive study by unprecedented numbers of observers around the world.  In the course of a day, the comet brightened from about 17th apparent magnitude up to naked-eye visibility, with concurrent expansion of an initially circular coma at the sky-plane velocity $\sim$550 m s$^{-1}$ (\citealp{2009AJ....138..625L,2010MNRAS...407.1784H}, corresponding to $\sim$40 arcseconds day$^{-1}$ at geocentric distance 1.6 AU).  

The remarkable photometric characteristics of 17P/Holmes introduced two practical problems for observers.  First, the high initial surface brightness of the coma caused saturation of the data from many instruments, especially those having large apertures and short focal ratios.  Second, the expansion of the coma soon over-filled the fields of view of many large telescopes, so that while photometry of the central regions could be obtained, photometry of the whole coma could not.  

In this paper, we report observations of 17P/Holmes fortuitously taken with the Solar Mass Ejection Imager (SMEI).  This orbiting instrument takes data with a 102 minute cadence well-suited to the study of the temporal evolution of the outbursting comet.  Moreover, SMEI images are obtained in such a way that even high surface brightness sources do not lead to saturation of the data, as we will describe.   Lastly, the angular resolution of SMEI permits measurements of the integrated light from the whole coma, at least for the first few months.   In these several regards, the SMEI data are complementary to other measurements taken with cameras that saturated \citep{2010MNRAS...407.1784H}, or which were unable to image the full coma owing to their limited fields of view \citep{2008A&A...479L..45M,2009AN....330..425M,2009AJ....138..625L}.  

%
%

\section{Observations} 
The Solar Mass Ejection Imager (SMEI) was launched on the Coriolis satellite by the United States Department of Defense in January 2003 \citep{2003SoPh..217..319E}.  The scientific aim of SMEI is to detect and forecast the arrivals of coronal mass ejections  (Jackson et al. 2004, Buffington et al. 2008).  SMEI has a Sun-synchronous polar orbit above the Earth's terminator, with a period of 102 minutes.  Three charge-coupled device (CCD) cameras, each with a field of view 60$^\circ \times 3^\circ$, scan the sky as the satellite orbits the Earth. They are oriented about $20^\circ$ above the local horizontal and pointed opposite to the motion of the spacecraft. Their alignments are such that Camera 1 points away from the Sun, Camera 3 points near the Sun and Camera 2 aims in the middle. This allows coverage of nearly the entire sky in one orbit.  The camera optics consist of two mirrors behind a complicated baffle structure, with an effective collecting area of $\sim 1.7$ cm$^2$.

The image scale of the camera is $0.05^\circ$ pixel$^{-1}$, but is degraded to $0.2^\circ$ pixel$^{-1}$ onboard during normal ``science mode'' operations.  In a normal astronomical camera system, such a large pixel scale would result in rapid saturation of the data from bright stars and even from high surface brightness coronal structures.  Two characteristics of the SMEI instruments prevent saturation of detector pixels caused by bright sources in the field of view. First, the exposure time for a single CCD frame is limited to only 4 seconds. Typically 1530 frames are combined from each CCD camera during a single orbit in order to produce one sky map. Second, the images from the camera are intentionally defocussed, such that point sources appear extended and fish-shaped in the plane of the CCD.  This reduces the likelihood of saturation by spreading the light from each point source over $\sim$200 pixels.  It also improves the photometric precision (up to 0.1\%) by allowing a large number of photoelectrons to be captured in each image without approaching the 350,000 electron full-well capacity of the CCD.  The instrument point spread function with a total width of $\sim 1^\circ$, provides a 0.1\% differential photometric sky brightness response as stellar signals sweep across the camera field of view.  The capacity to image bright sources without approaching saturation of the detector is a key advantage of SMEI when used to study outbursting comet 17P/Holmes.

Final images from SMEI are digitally constructed in J2000 equatorial coordinates. The data are re-sampled back to 0.1$\degr$ pixel$^{-1}$, to create sky maps for each SMEI camera with dimensions $3600\times 1200$ pixels in longitude and latitude \citep{2004SoPh..225..177J}.  The processing steps used at UCSD to convert the raw CCD images into photometrically accurate white-light sky maps include: integration of new data into the SMEI database; removal of an electronic offset (bias) and dark current pattern; identification of cosmic rays, space debris and Òflipper pixelsÓ (see \citealp{2005SPIE.5901..340H}, for further details); and placement of the images onto a high-resolution sidereal grid using spacecraft pointing information. To reduce background subtraction uncertainties, stars brighter than 6th magnitude are automatically removed from SMEI images by fitting the Point-Spread-Function \citep{2007SPIE.6689E...9H}.  

To avoid confusion between the multiple time-systems used to report observations (local time, Universal time, decimal Julian Day numbers, and modified Julian Day numbers have all been used), we employ the Day of Year (DOY) number, defined as being DOY = 1.0 on UT 2007 January 1 and increasing linearly thereafter (i.e. UT 2008 January 1 is DOY 366).  In this system, the perihelion of 17P/Holmes occurred on DOY = 124.6615, JD = 2454225.1615, UT 2007 May 04.6615.

\begin{figure}
\begin{center}
\includegraphics[width=1.0\textwidth]{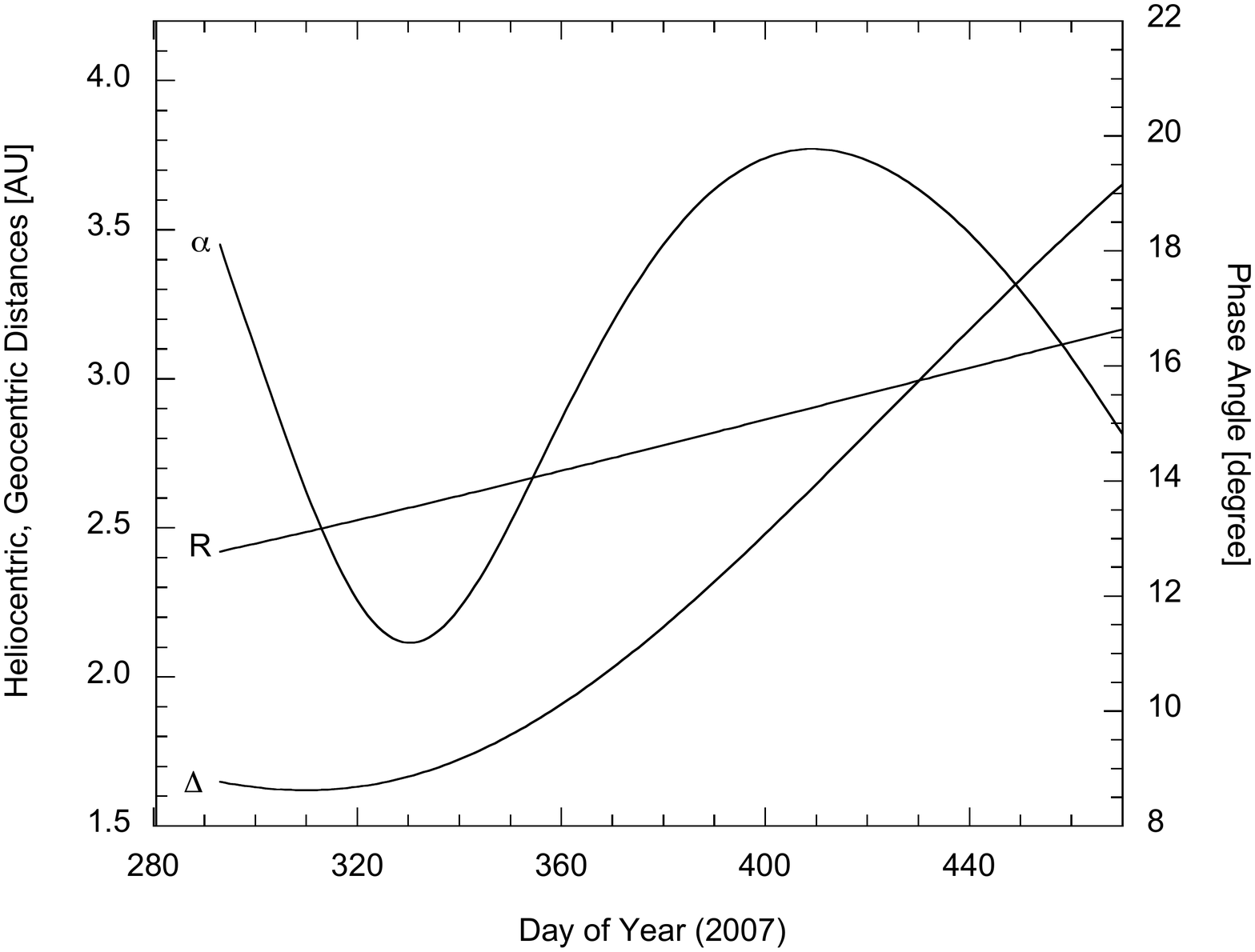}
\caption{(left axis) Heliocentric and geocentric  distances of 17P/Holmes, $R$ and $\Delta$ respectively,  and (right axis) phase angle, $\alpha$, as functions of time, expressed as day-of-year in 2007.\label{rda_plot}} 
\end{center} 
\end{figure}

The first SMEI sky map image showing 17P/Holmes has mid-integration time UT 2007 October 24 06h 37m 02s (DOY 297.275, the sky map was made between 05h 36m 12s and 07h 37m 52s) on Camera 1. Two previous images from the same day appear blank, apparently because the shutter of SMEI was closed.  The comet is already bright when first recorded and continued to be well-recorded by SMEI Camera 1 to  2008 January 11 08h 36m 26s (mid-time between 07h 35m 36s and 09h 37m 16s) and Camera 2  from UT 2008 January 1 01h 34m 56s (mid-time between 00h 43m 44s and 02h 26m 08s) to 2008 April 6 22h 15m 58s (mid-time between 21h 31m 24s and 23h 01m 32s). After April 6, the comet became too faint to be readily measured using SMEI. The observations covered a 165 day period with a total of 1992 sky map images. During this time, the geocentric distance doubled, while the heliocentric distance increased only slightly. The change in the observing geometry is illustrated in Figure \ref{rda_plot}.  Sample SMEI images of 17P/Holmes are shown in Figure \ref{images}.

\section{Analysis}

\subsection{Brightness Calibration}

Data from SMEI are routinely photometrically calibrated using bright stars distributed around the sky. However, because the passband of this filterless instrument is very broad and different from the standard astronomical filters, we elected to calibrate the data against our
own measurements of 17P/Holmes taken nearly simultaneously.  For these, we used the University of Hawaii 2.2-m telescope to image 17P/Holmes on UT 2007 October 26.33 (DOY=299.33) in order to photometrically calibrate the SMEI data.  The spectral response of the SMEI imager is very broad, exceeding 10\% in the optical wave band (4500-9500\AA) and $>$40\% over the 6000 $\le \lambda \le$ 7500\AA~wavelength range.  The central wavelength corresponds approximately to the astronomical $R$-band.  Accordingly,  an $R$-band filter was employed at the 2.2-m telescope and calibrated in the Kron-Cousins photometric system \citep{1992AJ....104..340L}.   The comet was imaged using a Tektronix 2048$\times$2048 pixel charge-coupled device camera placed at the f/10 Cassegrain focus, where the plate scale is 0.219$\arcsec$ per pixel and the field-of-view 450$\arcsec \times$450$\arcsec$.  We used aperture photometry with circular projected apertures and experimented to determine the optimum aperture radius for 17P/Holmes photometry.  We found that an aperture radius of 800 pixels (175$\arcsec$) was sufficient to capture $>$99\% of the light from the comet on this date.  Such a large aperture could not be used to measure the ($\sim$60,000 times fainter) standard star without incurring unacceptable errors from uncertainty in the sky background.  Instead, an aperture 20 pixels (4.4$\arcsec$) in radius, with sky determined from the median of data numbers in a surrounding annulus extending to 70 pixels (15.3$\arcsec$) radius, was used to measure the \citet{1992AJ....104..340L} standard star SA95-98. Again, we checked to be sure that this aperture captured $>$99\% of the light from the star.  

The particular circumstances of 17P/Holmes demand special mention here.   The high surface brightness of the coma on UT 2007 October 26 forced the use of unusually short integrations.  Normally, the Tektronix CCD camera is not used with exposures $<$5 s and, at shorter integration times, the linearity of the shutter (a spring-triggered leaf shutter)  is in question.  Spatial non-uniformity of the shutter  open time with position on the CCD degrades, as does knowledge of the exact duration of the open time.  To measure the importance of these effects we compared exposures of 0.1, 0.5, 1.0 and 5.0 seconds to estimate possible photometric errors arising from the forced use of short integrations on 17P/Holmes.  We find that systematic shutter errors are less than $\sim$10\% for the 17P/Holmes data.   This is small enough to be of no significance in the interpretation of the SMEI data.     

\subsection{Photometry of 17P/Holmes} 
\begin{figure}
\begin{center}
\includegraphics[width=0.9\textwidth]{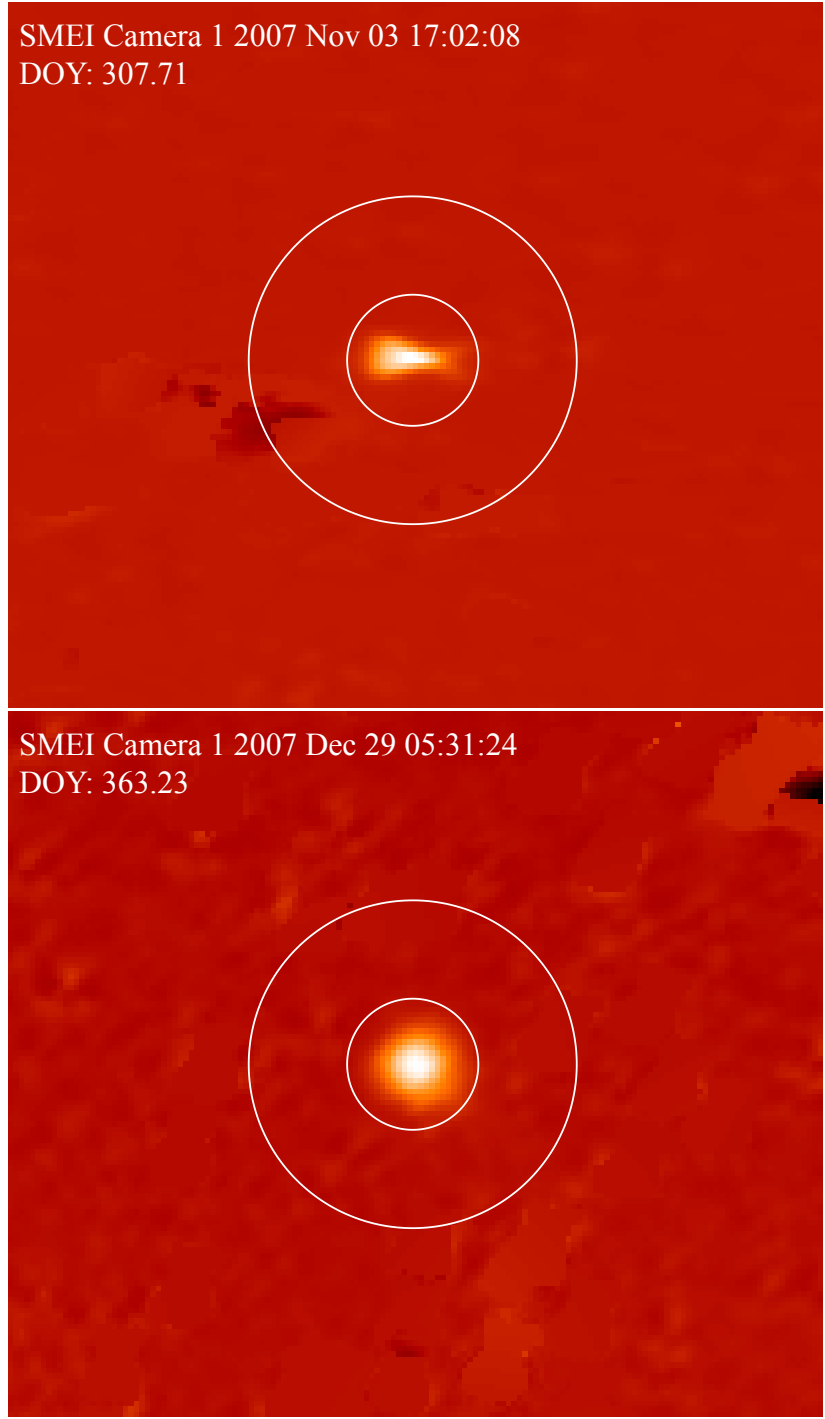}
\caption{Sample images of 17P/Holmes from SMEI Camera 1 taken 2007 November 3 (top) and December 29 (bottom). The region shown in each panel is $14.9\degr \times 12.9\degr$ ($149\times 129$ pixels) across with North to the top and East to the left.  Background stars brighter than 6th magnitude have been removed. The circles around comet 17P/Holmes have radii 1.2$\degr$, and 3.0$\degr$, respectively. On November 3, Holmes was unresolved, showing the intrinsic, fish-like SMEI image shape (top). By the end of December 2007, 17P/Holmes was partially resolved by SMEI so that the image appears more as a fuzzy ball (bottom). \label{images}} 
\end{center} 
\end{figure}

The brightness of 17P/Holmes was measured within projected, circular apertures centered on
the photocenter of the object.  Use of small apertures is precluded by the large point-spread-function
produced by SMEI, while large apertures suffer excessive contamination by background sources.  Accordingly, we employed a standard photometry aperture radius of 12 pixels (1.2$^\circ$) for our measurements, with sky subtraction determined from a contiguous annulus extending
to an outer radius of 30 pixels (3.0$^\circ$) (see Figure \ref{images}).  We used the median of the pixel values within the sky annulus
to define the sky brightness, since the median confers some protection against contamination of the
sky brightness by imperfectly removed field stars.  

\begin{figure}
\begin{center}
\includegraphics[width=0.9\textwidth]{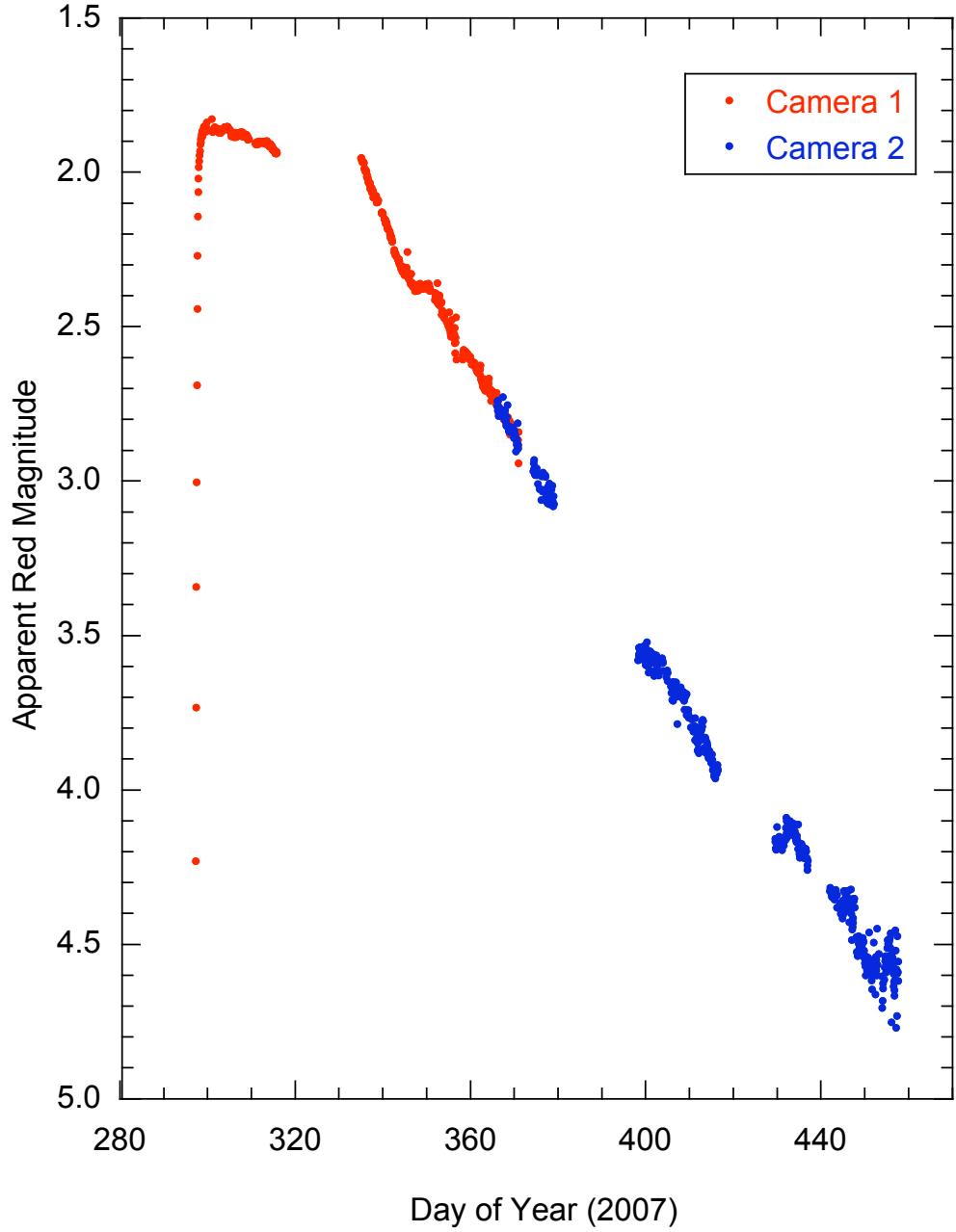}
\caption{Lightcurve of 17P/Holmes deduced from SMEI data. Gaps in the data show where bright field stars contributed excessive contamination.\label{mr_vs_doy1}} 
\end{center} 
\end{figure}

The photometry is shown in Figure \ref{mr_vs_doy1} as a function of 
time, with measurements from Cameras 1 and 2 identified.  Only Camera 1 measurements were calibrated against (nearly) simultaneous  observations from the University of Hawaii telescope.  However, the two SMEI cameras provide overlapping coverage in the period 366 $< DOY <$ 371, allowing us to calibrate Camera 2 against Camera 1.  Based on this overlap, we have normalized the photometry by subtracting 0.08 mag. from the Camera 2 measurements.  Gaps in the lightcurve in Figure \ref{mr_vs_doy1} appear where field stars have irreversibly compromised the comet data leading to their removal.  Remaining excursions in the lightcurve in Figure 3 (e.g. near DOY 340) result from residual contamination of the photometry by the wings of bright, distant stars.

To see the effects of viewing geometry on the lightcurve, we correct the apparent magnitudes, $R$, from Figure \ref{mr_vs_doy1}, to absolute magnitudes, $R(1,1,0)$, using

\begin{equation}
R(1,1,0) = R - 5 log_{10} (r \Delta) - 2.5 log_{10} \Phi(\alpha)
\end{equation}

\noindent in which $r$ and $\Delta$ are the heliocentric and geocentric distances, in AU, and $\Phi(\alpha)$ is the scattering phase function of the comet at phase angle $\alpha$.   The phase functions of active comets are difficult to measure because the phase changes are difficult to isolate from
simultaneous changes in $R$ and $\Delta$.  However, published phase functions are broadly consistent in
showing a large forward scattering peak and a more modest back-scattering peak \citep{1982AJ.....87.1310M,1987A&A...187..585M,1998Icar..132..397S}. For this work, we fitted the phase function of Schleicher et al. 1998 (for 0 $\le \alpha \le$ 70$^{o}$), to obtain

\begin{equation}
-2.5 log_{10} \Phi(\alpha) = 0.045 \alpha - 0.0004 \alpha^2.
\label{phasefunction}
\end{equation}

\noindent  Near opposition, Equation (\ref{phasefunction}) gives a phase coefficient of order 0.04 mag. degree$^{-1}$,
close to the characteristic values measured for the macroscopic surfaces of low albedo asteroids and cometary nuclei (Li et al. 2009, and references therein).  This suggests that the back-scattering properties of the dust are dominated by particles which are optically large ($2 \pi a/\lambda \ge$ 1, or $a \ge$ 0.1 $\mu$m, given $\lambda \sim$ 0.6 $\mu$m).  This, in turn, is compatible with the optical continuum colors, which are slightly redder than sunlight \citep{2009AJ....138..625L}, and with inferences from the coma of comet P/Halley, in which particles with 
$a <$ 0.1 $\mu$m were found to contribute negligibly to the integrated scattering cross-section \citep{1987A&A...187..767L}.  We note that the selection of the particular form of the phase function given by Equation (\ref{phasefunction}) is not critical, since the range of phase angles
over which 17P/Holmes was observed was modest (8.5$\degr$ $\le \alpha \le$ 19$\degr$) and the effects of phase in Figure \ref{mr_vs_doy1} are small compared to the effects of the varying geocentric distance. 

\begin{figure}
\begin{center}
\includegraphics[width=0.9\textwidth]{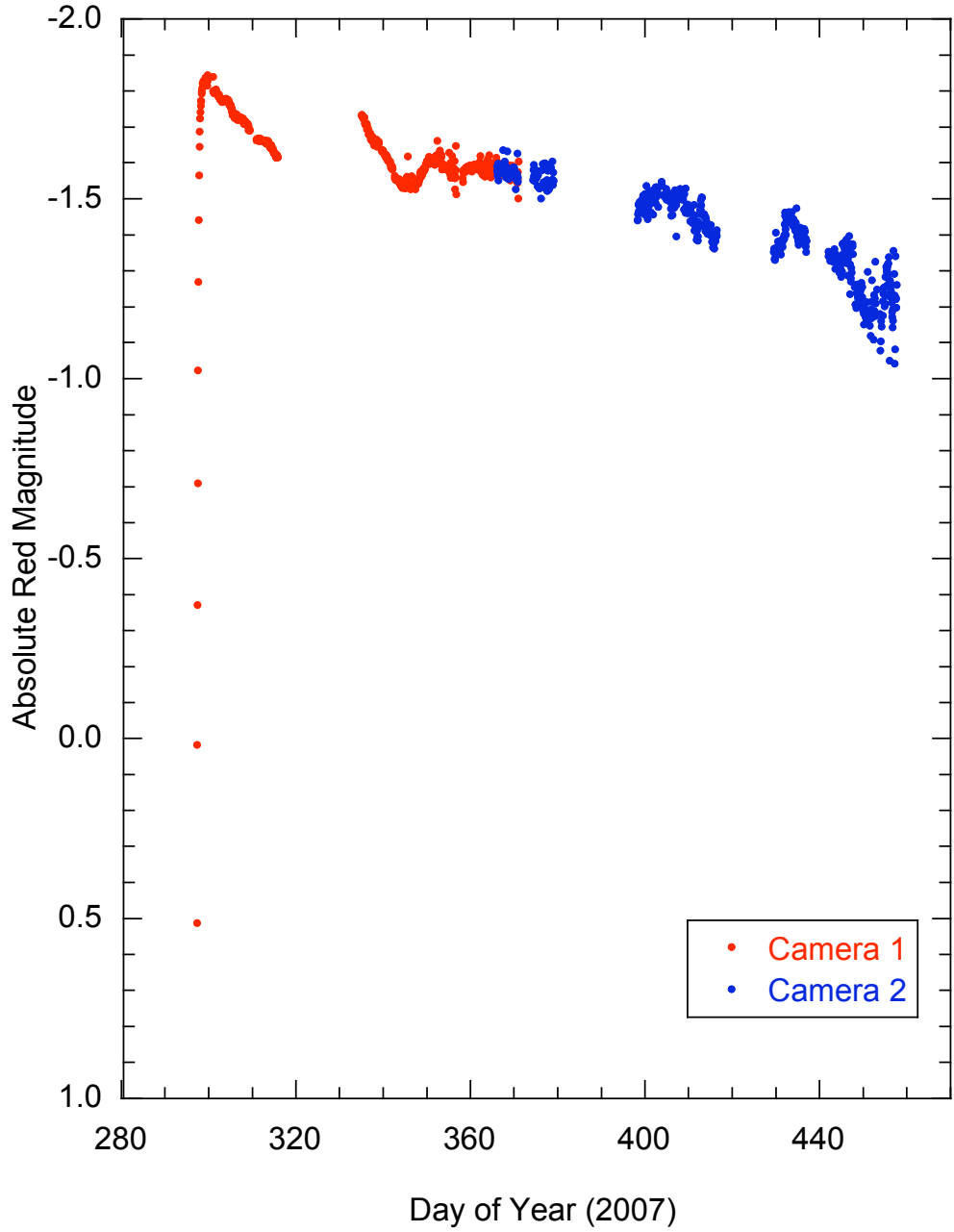}
\caption{The lightcurve of 17P/Holmes corrected for the effects of changing observing geometry and normalized to unit heliocentric and geocentric distances, and to zero phase angle. Gaps in the data show where bright field stars contributed excessive contamination.\label{mr_vs_doy2}} 
\end{center} 
\end{figure}

The resulting absolute magnitudes are shown in Figure \ref{mr_vs_doy2}.  Comparison with Figure \ref{mr_vs_doy1} reveals that the steep decline by about 2.5 mag. in apparent brightness observed for DOY $>$320 is largely a geometric artifact.  The fading portion of the lightcurve in Figure \ref{mr_vs_doy2} is comparatively gentle, dimming by only $\sim$0.6 mag.  over the same period.  This fading is dominated by the escape of dust particles from the region of the coma sampled by the photometry aperture.  Figure \ref{images} shows the change in appearance of the comet in SMEI images resulting from the partial resolution of the expanding coma by the end of 2007 December.  Evidence from other observers using smaller apertures confirms this conclusion.  For example, \citet{2009AN....330..425M} used small aperture photometry and found fading of the apparent magnitude by $\sim$8 mag. in the 100 days after the outburst whereas our integrated light photometry shows fading by $\sim$2 mag. over the same period (Figure \ref{mr_vs_doy1}), almost all of which is due to the changing observing geometry.

\section{Discussion}
\subsection{Scattering Cross-section}
\begin{figure}
\begin{center}
\includegraphics[width=0.9\textwidth]{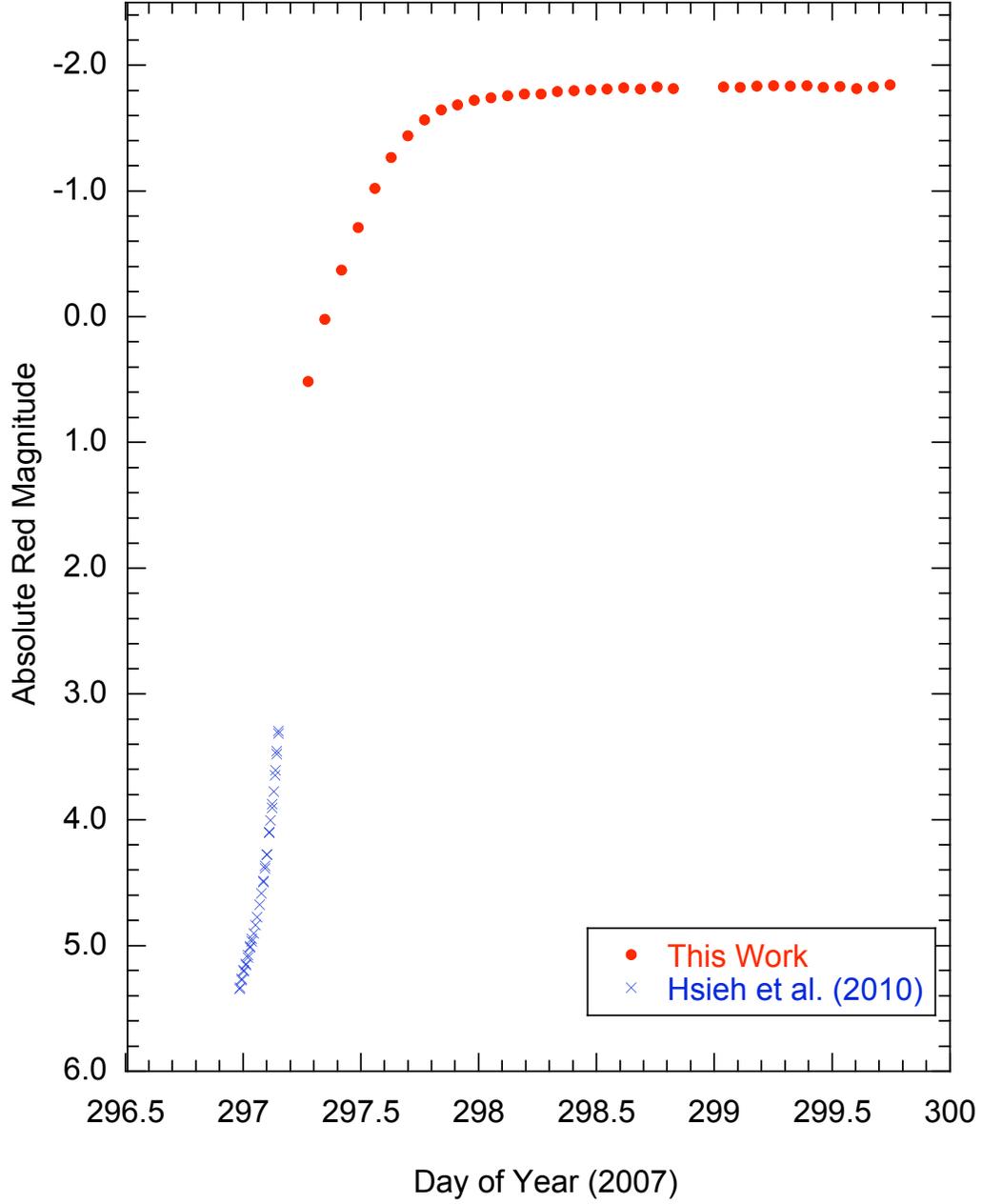}
\caption{Absolute lightcurve from SMEI data compared with data from SuperWASP \citep{2010MNRAS...407.1784H}.\label{comb_vs_DOY}} 
\end{center} 
\end{figure}

The absolute magnitudes can be used to measure the effective scattering cross-section of 17P/Holmes, $C_e$ (m$^2$), from 

\begin{equation}
p_R  C_e = 2.24 \times 10^{22}  \pi 10^{0.4(R_{\odot} - R(1,1,0))}
\label{square}
\end{equation}

\noindent in which $p_R$ is the geometric albedo measured in the $R$-band and $R_{\odot}$ = -27.11 is the apparent red
magnitude of the Sun \citep{1916ApJ....43..173R}.  We take the geometric albedo of cometary dust to be $p_R$ = 0.1 \citep{2002EM&P...90..497L,2009P&SS...57.1118H}.   Effective cross-sections computed in this way are plotted in Figure \ref{comb_vs_DOY}, where we also show photometry taken in the hours preceding the first SMEI observation, by \citet{2010MNRAS...407.1784H}.  The cross-sections are very large, rising to $C_e$ = 5.5$\times$10$^{13}$ m$^2$ by DOY= 
299.25, and equivalent to a circle of diameter 8.4$\times$10$^6$ m, considerably larger than the diameter of the Moon.   We note that the SMEI lightcurve in Figure \ref{comb_vs_DOY} is qualitatively similar to the lightcurve compiled by \citet{2008ICQ....30....3S} from visual and other data taken using a wide range of instruments and techniques.   However, the latter author derived a peak magnitude $H_0$ = -0.53$\pm$0.12 from naked eye observations, whereas SMEI data give $R(1,1,0)$ = -1.8$\pm$0.1 (see Fig. \ref{comb_vs_DOY}).  Part of the difference (perhaps $\sim$0.5 mag.) can be attributed to the continuum color of the comet \citep{2009AJ....138..625L} and the different effective wavelengths of the two measurements.  The remainder probably reflects difficulty in using the naked eye to measure the brightness of a diffuse but centrally-condensed source.  

\begin{figure}
\begin{center}
\includegraphics[width=1.0\textwidth]{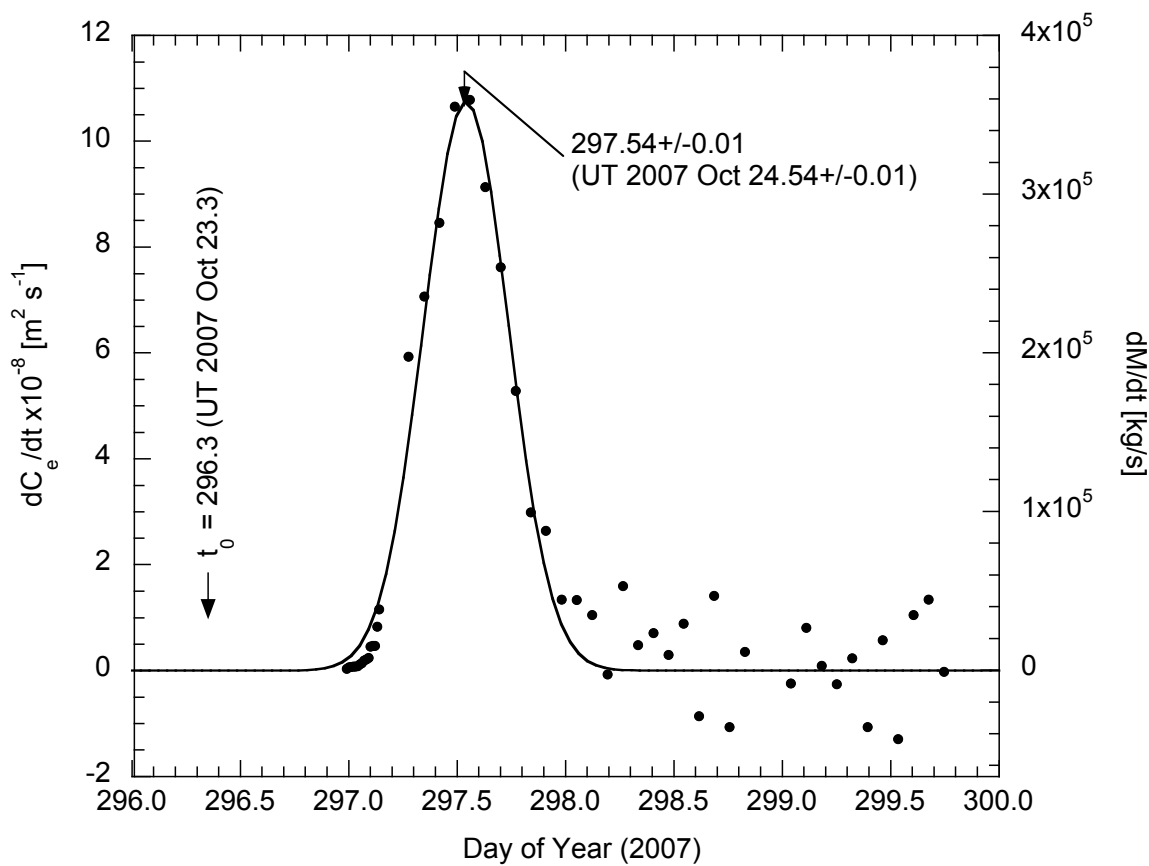}
\caption{(left) Rate of change of scattering cross-section and (right) rate of change of dust mass, as functions of time.  The solid line shows a Gaussian function fitted to the data.  The approximate time of the start of the outburst, $t_0$ = 296.3, is indicated.  The mass production rate refers to an effective particle size $a$ = 0.65 $\mu$m, as described in the text, and is an effective minimum.\label{dcbt_vs_DOY2}} 
\end{center} 
\end{figure}

The rate of change of the cross-section, $dC_e/dt$ (m$^2$ s$^{-1}$), is plotted in Figure \ref{dcbt_vs_DOY2}, for a 3-day period containing the start of the outburst.  Figure \ref{dcbt_vs_DOY2} shows that area production peaks at 11.0$\times$10$^8$ m$^2$ s$^{-1}$ on UT 2007 October 24.54$\pm$0.01 (DOY=297.54$\pm$0.01), about 0.5 day before the comet attains peak brightness (and cross-section), as seen in Figure \ref{comb_vs_DOY}.  The peak rate of brightening follows the estimated start of the outburst event (UT 2007 October 23.3$\pm$0.3, or DOY=296.3$\pm$0.3, \citealp{2010MNRAS...407.1784H}) by 1.2$\pm$0.3 days.

\subsection{Optical Depth}
To what extent is the lightcurve in Figures \ref{mr_vs_doy2}, \ref{comb_vs_DOY} and \ref{dcbt_vs_DOY2} influenced by optical depth effects in the coma?  The mean scattering optical depth is given by

\begin{equation}
\overline \tau(t) = \frac{C_e(t)}{\pi r_o(t)^2}
\label{tau}
\end{equation}

\noindent where $r_o(t)$ is the instantaneous radius of the coma.  We write $r_o(t)$ = $V (t - t_0)$, where $V$ = 550 m s$^{-1}$ is the expansion speed of the coma and $t_0$ = DOY 296.3 is the time of the start of the outburst \citep{2010MNRAS...407.1784H,2010Icar..208..276R}.  The spatially-averaged optical depth computed from Equations (\ref{square}) and  (\ref{tau}) is plotted in Figure (7), where we see that the peak value in the interval of observations, $\overline \tau$ = 3$\times$10$^{-3}$, was attained at DOY 297.8, about 1.5 days after the start of the outburst.  This does not rule out the possibility that the coma was globally optically thick before it was first observed.

\begin{figure}
\begin{center}
\includegraphics[width=0.9\textwidth]{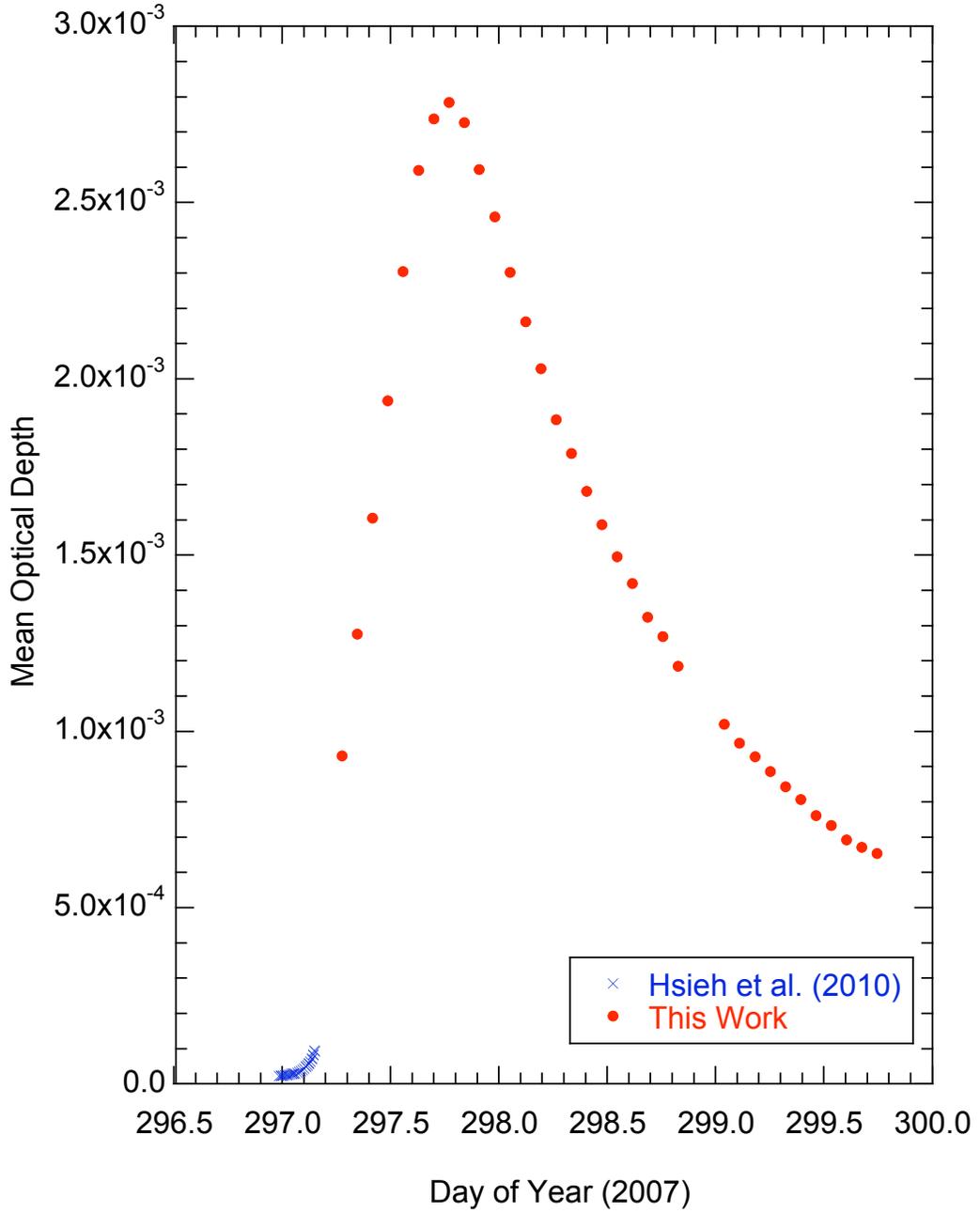}
\caption{Average optical depth versus time near the start of the outburst computed as described in the text. \label{tau_vs_DOY}} 
\end{center} 
\end{figure}

Figure \ref{tau_vs_DOY} shows that the coma was optically thin, \textit{on average}, even when at peak brightness, in agreement with the conclusion of \citet{2010MNRAS...407.1784H}.  Nevertheless, it is still possible that the coma was optically thick when measured along a line to the nucleus, a possibility  that we address here with a simple model.  As a reference point, we assume a spherically symmetric coma in which the number density of dust grains varies with the inverse square of the distance from the nucleus.  The line of sight optical depth, in the optically thin limit, then varies as $\tau(p) \propto p^{-1}$, where $p$ is the angle between a given line of sight through the coma and the direction to the nucleus.  We write

\begin{equation}
\tau(p) = \tau_n \left[\frac{p_n}{p}\right]
\label{taup}
\end{equation}

\noindent where $p_n$ = $r_n/\Delta$ is the angle subtended by the nucleus radius as observed from Earth and $\tau_n$ is the optical depth along a line to the center of the nucleus.  We assume $r_n$ = 1.7 km (Lamy et al. 2009) to find, at $\Delta$ = 1.6 AU, $p_n$ = 1.5$\times$10$^{-3}$ arcsec.  The average optical depth across the coma is

\begin{equation}
\overline \tau = \frac{\int_{p_n}^{p_o}2 \pi p \tau(p) dp}{\int_{p_n}^{p_o}2 \pi p dp}
\label{taubar}
\end{equation}

\noindent in which $p_o$ = $r_o(t)/\Delta$ is the angular radius of the coma at the instant when $\overline \tau$ is computed.  After substitution and rearrangement, Equations (\ref{taup}) and (\ref{taubar}) give

\begin{equation}
\tau_n = \frac{r_o(t)}{2 r_n} \overline \tau
\label{taun}
\end{equation}

\noindent provided $p_n \ll p_o$.  



\begin{figure}
\begin{center}
\includegraphics[width=0.9\textwidth]{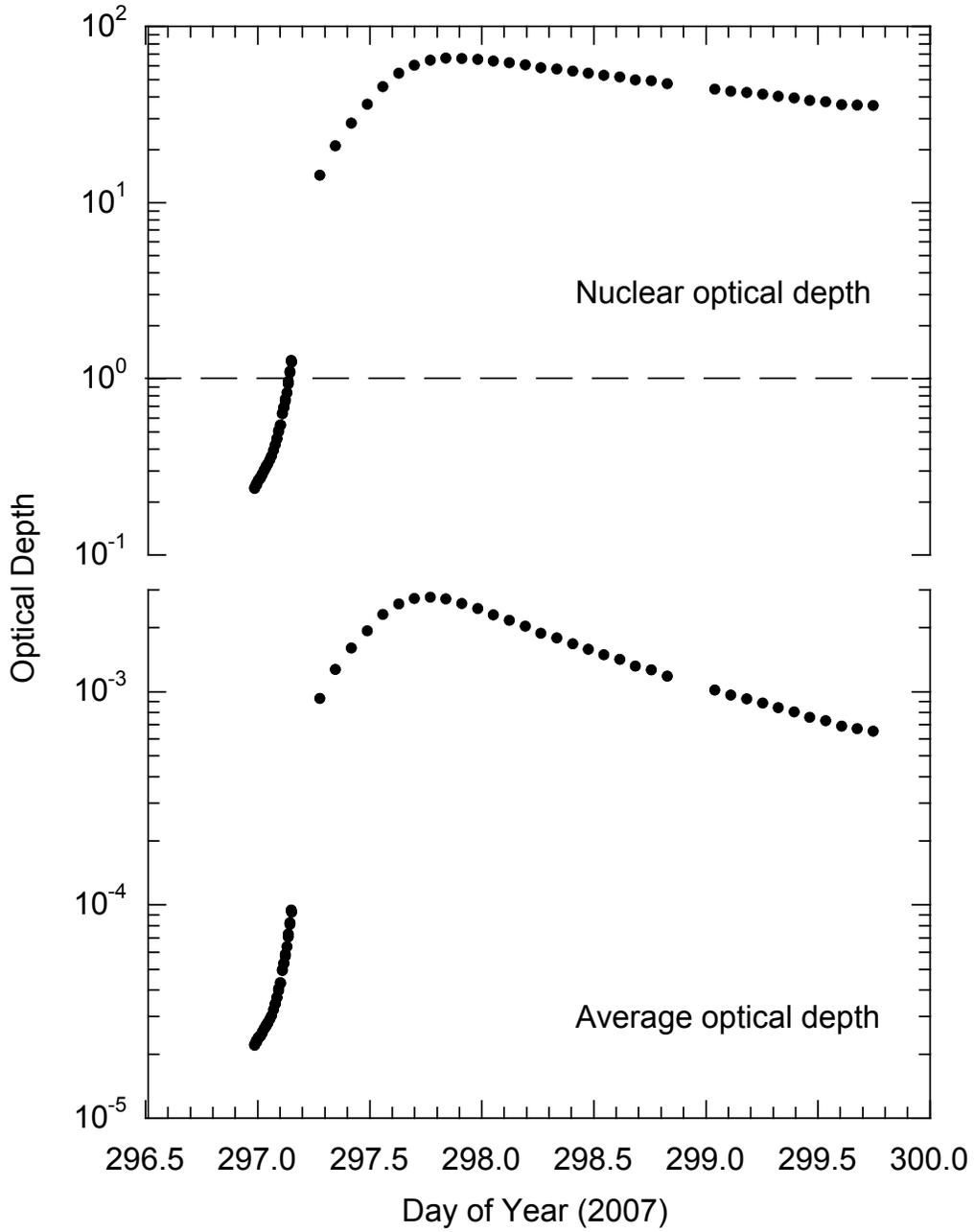}
\caption{Average and peak optical depth versus time. The vertical axis is broken for clarity of presentation, and a horizontal, dashed line shows the region where the coma is optically thick on a line to the center of the nucleus.\label{tau_vs_DOY2}} 
\end{center} 
\end{figure}

Equation (\ref{taun}) is a crude approximation in that it assumes spherical symmetry and a $p^{-1}$ coma.  Still, to order of magnitude, Equation (\ref{taun}) gives a useful estimate of the likely peak optical depths towards the nucleus. Figure \ref{tau_vs_DOY2} shows that, whereas the average $\overline \tau$ is always very small compared to unity, the coma may be optically thick along a line to the nucleus.    The nucleus received no direct sunlight as a result of the outburst, proving that the energy driving the expansion was either stored or derived from another source.

The projected angular radius of the optically thick region of the coma is obtained by setting $\tau(p)$ = 1 in Equation (\ref{taup}), giving $p$ = $\tau_n p_n$.  At the peak $\tau_n \sim$ 65 (DOY 297.8 from Figure 8), with $p_n$ = 0.0015$\arcsec$, we find that the optically thick region subtends an angle $p$ = 0.1$\arcsec$ as seen from Earth.  This is comparable to the ($\sim$0.06$\arcsec$) angular resolution offered by the best adaptive optics systems on large telescopes, or by the Hubble Space Telescope.  It is also so small that the integrated photometric characteristics of 17P/Holmes are dominated by scattering from the much larger optically thin region of the coma.  Our conclusion that the coma was optically thick only in a miniscule central region is supported by the optical observations from the UH 2.2-m telescope on UT 2007 October 26 (DOY 299) that were used to photometrically calibrate the SMEI DATA.  They show stars undimmed through the coma and a coma surface brightness increasing smoothly with decreasing projected distance from the nucleus.  On the other hand, a detection of extinction was reported by \citet{2008A&A...479L..45M} two days later, on UT 2007 October 28 (DOY 301). They observed the fading of stars, at  3$\sigma$ levels of confidence, separated from the nucleus by 25$\arcsec < p <$ 180$\arcsec$. Our data indicate immeasurably small central optical depths 3$\times$10$^{-4}< \tau <$ 2.5$\times$10$^{-3}$ at these large projected distances and thus cannot be reconciled with the observations of \citet{2008A&A...479L..45M}.

The above considerations show that optical depth effects play a negligible role in shaping the overall photometric properties of 17P/Holmes in outburst.  Instead, the lightcurve results from both the time dependence of the rate of release of mass (and cross-section) from the nucleus into the coma and the possible evolution of the scattering properties of particles once ejected. Near infrared spectral observations of the inner coma in late October and early November revealed water ice whose sublimation in sunlight would provide a natural mechanism for disaggregating composite grains (Yang et al.~2009).  Imaging observations show sub-structure suggestive of the breakup or fragmentation of centimeter- and decimeter-sized objects ejected from the nucleus of 17P/Holmes (Stevenson et al.~2010).   \citet{2010MNRAS...407.1784H} attempted to fit the early portion the lightcurve with a model assuming exponential fragmentation of dust particles and obtained fits with decay timescales of 1000 s and 2000 s.  However, fits to data from their limited ($\sim$4 hr) observing window do not match the more extensive SMEI data-set presented here, and any simple model of the lightcurve in terms of dust fragmentation cannot be supported.  All we can say based on the lightcurve is that the brightening reflects the combined effects of the time-dependent nucleus mass production function (assumed to be impulsive by \citealp{2010MNRAS...407.1784H}) and evolutionary changes in the dust scattering properties.  The data offer no way to separate these two effects.

\subsection{Mass and Energy}
The conversion between the derived scattering cross-section and the particle mass is model-dependent and very uncertain.  The principal unknown is the dust size distribution, but the particle density is also unmeasured and its value must be assumed.  The simplest model is to assume that the particles are all spheres of one effective radius, $a_e$, and density, $\rho$.  Then, the total dust mass is given by

\begin{equation}
M = \frac{4}{3} \rho a_e C_e.
\label{masss}
\end{equation}

\noindent Solid spheres scatter electromagnetic radiation most efficiently when $a \sim \lambda$ \citep{1983B}.  With $a_e = \lambda$ = 0.65 $\mu$m and $\rho$ = 400 kg m$^{-3}$, the peak $C_e$ = 5.5$\times$10$^{13}$ m$^2$ (Figure (5)) gives mass $M$ = 1.9$\times$10$^{10}$ kg.  The mass of the nucleus,  taken to be a sphere of radius 1.7 km and having the same density, is $M_n$ = 8$\times$10$^{12}$ kg, so that $M/M_n \sim$ 0.2\%.

However, this simplest case is likely to underestimate the dust mass, because the real particles will occupy a size distribution in which large particles might contain significant mass while presenting negligible cross-section.  Optical data alone provide little or no evidence concerning such particles but we can estimate an upper limit to the dust mass as follows. The spectral energy distribution from optical (0.5 $\mu$m) to mid-infrared (20 $\mu$m) wavelengths has been modelled in 17P/Holmes in terms of power-law distributions of dust particles size in which the number of particles having radius between $a$ and $a + da$ is proportional to $a^{-q} da$ (Ishiguro et al. 2010).  The models indicate $q >$  3 over the radius range 0.3 $\mu$m to 100 $\mu$m.   Measurements in other disintegrating comets show that, while the size distribution is not precisely described by a power law of any index, the data are broadly compatible with power law models 3 $< q < $ 4 \citep{2007PASJ...59..381F,2010Nature,2010AJ....139.1491V}.  We consider a middle value, $q$ = 3.5, with minimum and maximum particle radii, $a_1$ and $a_2$, respectively.  The effective radius is then $a_e = (a_1 a_2)^{1/2}$.  With $a_1$ = 0.1 $\mu$m (particles much smaller than this have negligible interaction with optical photons and so present no cross-section for scattering) and $a_2$ = 10$^{-2}$ m (Gr{\"u}n et al. 2001), we obtain $a_e = (a_1 a_2)^{1/2}$ = 30 $\mu$m.  The mass computed from Equation (\ref{masss}) then rises to $M$ = 9$\times$10$^{11}$ kg, or $M/M_N \sim$ 10\%.  The range of inferred dust masses, 2$\times$10$^{10}$  $< M <$ 90$\times$10$^{10}$ kg, may be compared with the best estimate from mid-infrared thermal observations, namely $M$ = 1.0$\times$10$^{10}$ kg (Reach et al. 2010). 

Masses near the upper limit, 2$\times$10$^{11}$ kg, have been claimed based on millimeter wavelength radio-continuum measurements \citep{2009A&A...495..975A}.  We have reinterpreted these measurements according to the formalism described in \cite{1992Icar..100..187J}. The principal uncertainty is the opacity.  With $\lambda$ = 1 mm opacities in the range 1 to 10 m$^2$ kg$^{-1}$, we estimate dust masses from the radio-continuum in the range 10$^9$ to 10$^{10}$ kg on UT 2007 October 27.1 within a 5.7$\arcsec \times$7.3$\arcsec$ beam.  These masses are 1 to 2 orders of magnitude smaller than derived by Altenhoff et al. (2009), but consistent with a re-analysis of the same radio-continuum data by Reach et al. (2010) and with the range of masses allowed by the SMEI photometry alone.  Unfortunately, even the earliest reported radio-continuum measurements sample only a tiny central region in the expanding coma.  For example, the first radio-continuum measurement on UT Oct 27.105 used an elliptical 5.7$\arcsec \times$7.3$\arcsec$ beam at a time ($\sim$4 days after the outburst) when the angular diameter of the dust coma was already 300$\arcsec$.  Therefore, the radio-continuum data on 17P/Holmes offer only a lower limit to the total dust mass.


For the rest of the discussion, we use  2$\times$10$^{10}$ $< M <$ 90$\times$10$^{10}$ kg as the best estimate of the dust mass.  
From Equation (\ref{masss}) the rate of dust production by mass is $dM/dt$ = $4/3 \rho a_e (dC_e/dt)$, giving $dM/dt \sim$ (3 to 140)$\times$10$^5$ kg s$^{-1}$ at the maximum on UT 2007 October 24.54$\pm$0.01 (DOY 297.54$\pm 0.01$).  No contemporaneous measurements of the gas production rate are available. The earliest reported gas production rate is by Combi et al. (2007), who measured $Q(H_2O)$ = 1.4$\times$10$^{30}$ s$^{-1}$ on UT 2007 October 27 (DOY 300), corresponding to 0.4$\times$10$^5$ kg s$^{-1}$.  Some 4 days after the start of the outburst and 3 days past its peak, the dust production at this time was already negligible (Figure 6).  \citet{2009AJ....138.1062S} extrapolated narrowband photometry data to infer peak water production rates $Q(H_2O) \sim$7$\times$10$^{29}$ s$^{-1}$, or 0.2$\times$10$^5$ kg s$^{-1}$.    

\subsection{Ejecta and Outburst Trigger}
The ejected dust mass is equivalent to a cube having a side length $(M/\rho)^{1/3}$ = 370 m to 1300 m, where $\rho$ = 400 kg m$^{-3}$ is the assumed bulk density of the nucleus.  However, this is an unlikely description of the outburst geometry, for three reasons.  Firstly, the sky-plane morphology of 17P/Holmes was initially circularly symmetric, with global deviations from circularity only appearing on timescales of a week after radiation pressure had begun to deflect coma dust \citep{2010AJ....139.2230S}.  Eruption of material from a localized surface source would more naturally produce a jet or a cone, not a spherical debris cloud or one that appeared symmetric in projection onto the sky. (It is sometimes argued that projection effects would hide deviations from circular symmetry because of the small phase angles of observation but, in fact, with an average phase angle $\sim$0.2 radian, (see Figure 1) any strong asymmetries would easily have been detected if present).  Secondly, the collimated ejection of mass from a spatially localized source would impart significant recoil to the motion of the nucleus.   Very roughly, the velocity impulse on the nucleus is given by $\Delta V$ = $(M/M_N) V$, where $V$ = 550 m s$^{-1}$ is the ejecta velocity.  Substituting 0.2\% $< M/M_N <$ 10\% gives $\Delta V$ = 1.5 to 70 m s$^{-1}$.  In one month, an impulse of this magnitude would lead to a displacement of the nucleus from its pre-outburst predicted position by 3900 km to 180,000 km, and this could scarcely have escaped detection.  (At our request, Dr. Brian Marsden examined the reported positions of 17P/Holmes before and after the outburst to search for evidence of a change in the fitted non-gravitational parameters, but found none).  Lastly, on physical grounds it is difficult to imagine a process that would drive mass-loss many hundreds of meters deep into the nucleus against the expected radial gradient of temperature from the hot surface to the cold interior.  

At the other extreme, the ejected mass could be contained in a surface layer on the nucleus having thickness 

\begin{equation}
\ell = \frac{M}{4\pi r_n^2 f \rho},
\label{ell}
\end{equation}

\noindent  where $f$ is the fraction of the surface area of the nucleus that is ejected.  Substituting $f$ = 1 gives 1.4 m $< \ell <$ 60 m.  Shell-like models have been championed for comets including 17P/Holmes for many years \citep{1982come.coll..251S,2008ICQ....30....3S,2009ICQ....31...99S}.  In these models, the rapid increase in brightness and scattering cross-section would be caused by disaggregation of the shell, presumably driven by sublimation of ices acting as glue in aggregated structures when freshly exposed to solar radiation and by collisions between disaggregated pieces moving at different speeds under gas drag near the nucleus.  Sekanina's model is not contradicted by any aspect of the SMEI photometry.  In this scenario, the 1.2$\pm$0.3 day lag between the start of the outburst and the peak rate of area production (Figure (6)) provides a measure of the timescale of the disaggregation.  

We compare  $\ell$ with the distance over which heat can be transported in the nucleus by conduction.  From solution of the heat diffusion equation, this distance is $\delta r = (\kappa P/\pi)^{1/2}$, where $\kappa$ is the thermal diffusivity of the surface materials and $P$ is the period of time over which conduction acts.  The thermal diffusivity of porous dielectrics is roughly $\kappa \sim$ 10$^{-7}$ m$^2$ s$^{-1}$.  For example, setting $P$ = 6.88 yr, the orbital period of 17P/Holmes,  we find, $\delta r$ = 2.5 m. Periodic forcing of the insolation as the nucleus moves around its eccentric orbit drives a wave of conducted heat into the nucleus that damps over a length scale $\delta r \sim$ 2.5 m.  In the $\sim$100 yrs elapsed since the outbursts of 1892/93, conducted heat would reach $\delta r \sim$ 25 m beneath the initial surface.  Regions with depth  $\gg \delta r$ will be largely immune to surface heating effects driven by recent surface events and thus are candidate locations for the survival of amorphous and trapped supervolatile ices.  An important conclusion is that, within (considerable) uncertainties, $\ell \sim \delta r$.  This approximate equality suggests that the outburst of 17P/Holmes in 2007 could have been triggered by heat conducted from the surface first exposed to space and direct sunlight by the outbursts of 1892/93.  

A plausible trigger is the crystallization of amorphous water ice, which is exothermic and which is expected to result in the release of trapped supervolatile gases capable of driving the outburst \citep{2004come.book..359P,2007Icar..190..655B,2010Icar..208..276R}.  Amorphous ice has not been directly detected, but provides a self-consistent explanation for the activity observed in comets \citep{2009Icar..201..719M} and Centaurs \citep{2009AJ....137.4296J} located beyond the orbit of Jupiter (where temperatures are too low for crystalline water ice to sublimate).  Crystallization models of comets necessarily assume values for many unknown or poorly-constrained physical parameters (e.g. the thermal diffusivity, the ice/rock ratio, the nucleus spin properties, the mass of trapped gas, even the orbital evolution in the recent past is important in determining the subsurface temperature structure). As a consequence, crystallization models are very flexible but also very difficult to reject based on observations.  One feature that is largely independent of the many unknowns is the step-wise progression of the crystallization front into the nucleus.  Thermal runaways triggered by crystallization near the surface propagate downward into colder ice. Eventually, the heat released by crystallization is insufficient to drive additional ice to crystallize, and the runaway stops.  The vertical distance is related to the thermal skin depth impressed in the nucleus by sunlight added at the surface and typically measured in meters.  Crystallization is therefore at least qualitatively consistent with a scenario in which a disintegrating dusty surface shell is launched from the 17P/Holmes nucleus.  

\citet{2010Icar..207..320K} computed thermal models of 17P/Holmes and reached the opposite conclusion, namely that runaway crystallization is unlikely to have been responsible for the outburst.  However, their conclusion relied, in part,  on the very high ejected mass estimates of 10$^{12}$ to 10$^{14}$ kg by \citet{2008A&A...479L..45M}.  As noted earlier, the latter mass estimates are based on a reported detection of extinction in the coma which sits uncomfortably with the large-aperture SMEI data presented here.   In fact, the upper end of the \citet{2008A&A...479L..45M} mass estimate considerably exceeds our best guess as to the mass of the entire nucleus of 17P/Holmes, and therefore cannot be correct.  For this reason, and because our own mass estimates (see also \citealp{2008ICQ....30....3S,2010ApJ...714.1324I,2010Icar..208..276R}) are considerably smaller, we consider that to reject crystallization as the energy source for the 17P/Holmes outburst would be premature.   



 If the particles all travel with characteristic speed $V$ = 550 m s$^{-1}$, their total kinetic energy is $E$ = (3 to 140)$\times$10$^{15}$ J, equivalent to about 0.7 to 33 Megaton of TNT (1 MT = 4.2$\times$10$^9$ J).  This, in turn, is equal to the total solar energy falling on the 1.7 km radius nucleus in 7 to 350 days.  While sunlight might be needed to trigger the outburst of 17P/Holmes, it clearly cannot supply enough energy to drive it.  The crystallization of amorphous water ice releases $\Delta E \sim$ 9$\times$10$^4$ J kg$^{-1}$.  Curiously, this is close to the energy per unit mass of the 17P/Holmes coma, $E/M \sim$ 10$^5$ J kg$^{-1}$.  Therefore, crystallization of a subsurface layer of amorphous ice with the associated release of trapped supervolatile gases  could supply the mass, energy and momentum of the ejecta responsible for the remarkable outburst of comet 17P/Holmes.  However, why 17P/Holmes should be uniquely afflicted by three such extraordinary outbursts, whereas most other comets show none, remains a complete mystery.

\clearpage

\section{Summary}

We have used photometric time-series data from the orbiting Solar Mass Ejection Imager to study the evolution of outbursting comet 17P/Holmes.  The SMEI's large pixel size and the broad field of view allow the spatially and photometrically full coverage of the comet during its rapid expansion due to outburst.

\begin{enumerate}

\item The comet was first detected by SMEI on UT 2007 October 24.275 (DOY 297.275) at apparent red magnitude 4.25, quickly brightened to 
peak brightness (apparent red magnitude 1.8) over the following day and thereafter faded over the next 5 months.

\item The coma remained globally optically thin (average optical depth  $<$3$\times$10$^{-3}$) at all times but is inferred to have been locally optically thick (on a line of sight to the nucleus) through the period of observations.

\item The mass of the dust coma was (2 to 90)$\times$10$^{10}$ kg, corresponding to 0.2\% to 10\% of the nucleus mass.  The ejected mass is equivalent to that contained within a surface shell on the 1.7 km radius nucleus having a thickness 1.4 m to 60 m.  Comparison with the $\sim$25 m thermal skin depth for heat conducted inwards since the previous outbursts in 1892/93, is consistent with conducted heat being the trigger responsible for the outbursts.

\item The rate of change of the scattering cross-section can be approximately matched by a Gaussian function having mid-time UT 2007 October 24.54$\pm$0.01 (DOY 297.54$\pm$0.01) and full-width at half-maximum 0.44$\pm$0.02 days.  Thus, there is a 1.2$\pm$0.3 day lag between the start of the outburst (as inferred from observations by \citealp{2010MNRAS...407.1784H}) and the time of peak activity that may measure the timescale for the disintegration of fragments in the coma.  Dust cross-section was added to the coma at a peak rate of 1070$\pm$ 40 km$^2$ s$^{-1}$.  

\item The kinetic energy of the outburst was in the range (3 to 140)$\times$10$^{15}$ J, far too large for sunlight to play any more than a triggering role in the expansion of the ejecta.   The energy per unit mass of the ejecta (10$^5$ J kg$^{-1}$) is of the same order as the energy per unit mass released upon the crystallization of amorphous water ice.

\end{enumerate}

\acknowledgments
The comet Holmes observation by SMEI was first brought to our attention by the web site: http://www.smei.nso.edu/gallery.html sponsored by Air Force Research Laboratory, Space Weather Center of Excellence, National Solar Observatory. We thank SMEI team member Dr. Pierre Hick for his generous contribution to the data processing description and Dr. Andrew Buffington for helpful discussions. We thank the referee and Michal Drahus for comments which helped to improve the presentation. SMEI was designed and constructed by a team of scientists and engineers from the US Air Force Research Laboratory, the University of California at San Diego, Boston College, Boston University, and the University of Birmingham, UK. This work was supported, in part, by grants to DJ from NASA's Planetary Astronomy and Outer Planets Research programs.

\clearpage

\clearpage

\clearpage

\clearpage

\clearpage

\clearpage

\clearpage

\clearpage

\end{document}